\documentclass[]{article}

\usepackage{times,latexsym}
\usepackage{url}
\usepackage[T1]{fontenc}
\usepackage{amssymb}
\usepackage{pifont}
\usepackage{graphicx}
\usepackage{adjustbox}
\usepackage{appendix}
\usepackage{natbib}
\usepackage{arxiv}
\usepackage{hyperref}
\usepackage{url,hyperref,lineno,microtype,subcaption}
\usepackage[onehalfspacing]{setspace}
\usepackage{float}
\usepackage{multirow}

\title{Enhancing AI microscopy for foodborne bacterial classification via adversarial domain adaptation across optical and biological variability}

\usepackage{authblk}

\author[2]{Siddhartha Bhattacharya}
\author[2]{Aarham Wasit}
\author[3,4]{Mason Earles}
\author[3,5]{Nitin Nitin}
\author[5]{Luyao Ma}
\author[1]{Jiyoon Yi\thanks{\texttt{yijiyoon@msu.edu}}}

\affil[1]{Department of Biosystems and Agricultural Engineering, Michigan State University, East Lansing, MI, USA}
\affil[2]{Department of Computer Science and Engineering, Michigan State University, East Lansing, MI, USA}
\affil[3]{Department of Biological and Agricultural Engineering, University of California, Davis, CA, USA}
\affil[4]{Department of Viticulture and Enology, University of California, Davis, CA, USA}
\affil[5]{Department of Food Science and Technology, University of California, Davis, CA, USA}

\date{}

\begin{document}

\newcommand{\xmark}{\ding{55}}%

\maketitle

\begin{abstract}
Rapid detection of foodborne bacteria is critical for food safety and quality, yet traditional culture-based methods require extended incubation and specialized sample preparation. This study addresses these challenges by i) enhancing the generalizability of AI-enabled microscopy for bacterial classification using adversarial domain adaptation and ii) comparing the performance of single-target and multi-domain adaptation. Three Gram-positive (\textit{Bacillus coagulans}, \textit{Bacillus subtilis}, \textit{Listeria innocua}) and three Gram-negative (\textit{E. coli}, \textit{Salmonella} Enteritidis, \textit{Salmonella} Typhimurium) strains were classified. EfficientNetV2 served as the backbone architecture, leveraging fine-grained feature extraction for small targets. Few-shot learning enabled scalability, with domain-adversarial neural networks (DANNs) addressing single target domains and multi-DANN (MDANNs) generalizing across all target domains. The model was trained on source domain data collected under controlled conditions (phase contrast microscopy, 60× magnification, 3-h bacterial incubation) and evaluated on target domains with variations in microscopy modality (brightfield, BF), magnification (20×), and extended incubation to compensate for lower resolution (20×–5h). DANNs improved target domain classification accuracy by up to 54.45\% (20×), 43.33\% (20×–5h), and 31.67\% (BF), with minimal source domain degradation (<4.44\%). MDANNs achieved superior performance in the BF domain and substantial gains in the 20× domain. Grad-CAM and t-SNE visualizations validated the model’s ability to learn domain-invariant features across diverse conditions. This study presents a scalable and adaptable framework for bacterial classification, reducing reliance on extensive sample preparation and enabling application in decentralized and resource-limited environments.
\end{abstract}

\section{Introduction}
\label{sec:section1}

Rapid detection and identification of foodborne bacteria, including both pathogenic and spoilage species, are essential for ensuring food safety and quality. However, traditional methods rely on time-intensive culture-based approaches, requiring prolonged incubation to form visible bacterial colonies \citep{Ferone2020}. These methods are further constrained by the need for selective media tailored to specific pathogens, limiting their scope and generalizability \citep{Ferone2020, Qiu2021}. Such challenges not only delay the implementation of corrective actions but also increase the risk of outbreaks, product recalls, and economic losses across the food supply chain \citep{Qiu2021, Hoffman2024}. Addressing these limitations requires innovative approaches that combine speed, accuracy, scalability, and minimized resource demands in bacterial classification.

Artificial intelligence (AI)-enabled microscopy has emerged as a promising solution, integrating deep learning for rapid analysis of microscopic patterns captured through quick imaging snapshots. Our previous work demonstrated that convolutional neural networks (CNNs) can classify bacteria at the microcolony stage, significantly reducing testing time compared to traditional methods that require lengthy enrichment and full incubation \citep{Ma2023}. However, this approach relied on an existing CNN architecture optimized for computational performance on generic datasets rather than addressing the unique challenges of microscopic imaging. Moreover, model training was conducted on datasets collected under controlled laboratory conditions, limiting its generalizability to real-world scenarios characterized by optical and biological variability. Similarly, other early efforts in AI-enabled microscopy relied on small datasets and generic architectures, increasing the risk of overfitting and restricted their applicability to diverse imaging conditions \citep{Melanthota2022, Wu2023}. These constraints highlight the critical need for tailored architectural advancements and methods that enhance robustness to variability.

Recent advancements in CNN architecture design and data-centric techniques have shown potential to address these challenges. EfficientNet, for example, employs a compound scaling approach that systematically balances depth, width, and resolution, enabling the capture of fine-grained morphological features in small targets while minimizing computational cost \citep{Tan2019}. EfficientNet variants have been applied to identify cellular defects and detect blood cells in hematology, showing their adaptability to small targets within cell imaging contexts \citep{Otamendi2021, Xu2022}. The newer EfficientNetV2 incorporates faster training and enhanced regularization, making it even more suitable for tasks requiring subtle or fine-grained feature extraction on limited datasets \citep{Tan2021}. Additionally, image augmentation techniques have proven effective in improving bacterial classification and generalization across growth stages in clinical microscopy by simulating variability inherent to imaging conditions \citep{Chin2024, Jeckel2020}. Together, these architectural innovations and data augmentation strategies address some limitations of previous approaches, enabling more effective analysis of small bacterial targets.

Beyond architectural advancements, domain adaptation techniques have been increasingly explored to address variability in optical setups and biological conditions \citep{Tomczak2021}. Traditional deep learning models often assume that training and testing data follow the same probability distribution. However, real-world applications like bacterial image classification encounter significant distribution shifts due to variations in microscopes, imaging conditions, or bacterial growth conditions across laboratories and testing sites. Collecting and annotating large datasets for every new environment is both impractical and expensive for biological images. Domain adaptation addresses this challenge by enhancing model robustness, enabling generalization to diverse setups without extensive data annotation. Previous studies applying domain adaptation in biomedical imaging have shown enhanced generalization across imaging modalities and biological conditions for tasks such as image classification and cell segmentation \citep{Tomczak2021, Xing2021}. Such techniques have shown promise for addressing variability while reducing reliance on annotated datasets.

Thus, this study aims to develop a domain-adaptive image classification model that generalizes from controlled laboratory conditions (\textit{source domain}) to variable laboratory conditions (\textit{target domains}), with only a few labeled examples. The objectives are to i) enhance the generalizability of AI-enabled microscopy for bacterial classification using adversarial domain adaptation and image augmentation, and ii) compare the performance of classical single-target domain adaptation with multi-domain adaptation for simultaneous generalization across multiple domains. To achieve these goals, domain-adversarial neural networks (DANNs) and multi-domain adversarial neural networks (MDANNs) were employed to learn shared, domain-invariant feature representations optimized through adversarial training. By implementing these advancements with as few as 1-5 labeled samples per bacterial species, this study effectively addresses challenges posed by natural biological variability and domain shifts in microscopic imaging.

\section{Materials and Methods}
\label{sec:section2}

\subsection{Source and target domain conditions}
\label{sec:section2.1}

In domain adaptation, the \textit{source domain} is the dataset used for training, collected under controlled conditions, while \textit{target domains} are datasets with variations used to test the model's generalizability. The source domain (`PC') dataset, primarily derived from our previous work \citep{Ma2023}, was collected under controlled laboratory conditions using phase contrast microscopy at 60× magnification with microcolonies incubated for 3 h. These conditions were identified as optimal for bacterial microcolony imaging in our prior study. To evaluate generalizability in this study, additional datasets were collected under varying conditions simulating optical setups and microbial variability. These target domains included: i) brightfield microscopy at 60× magnification (`BF' domain), representing lower-contrast imaging often used in resource-limited setups; ii) phase contrast microscopy at a lower magnification of 20× (i.e., `20×' domain), using more accessible, less specialized equipment that trades resolution for broader applicability; and iii) phase contrast microscopy at a 20× magnification with an extended incubation time of 5 h (i.e., `20×–5h' domain), capturing the additional biological variability introduced by longer growth periods, which might arise from deviations in protocol or the need to enhance detectability with lower magnification. A summary of laboratory conditions for all domains is presented in Table \ref{tab:table1}. 

\begin{table}[h]
\centering
\caption{Source and target domains with laboratory conditions}
\begin{tabular}{lcccc}
\hline
\textbf{Abbreviation} & \textbf{Domain type} & \textbf{Microscopy modality} & \textbf{Magnification} & \textbf{Incubation time} \\
\hline
PC & Source & Phase contrast & 60× & 3 h \\
BF & Target & Brightfield & 60× & 3 h \\ 
20× & Target & Phase contrast & 20× & 3 h \\
20×–5h & Target & Phase contrast & 20× & 5 h \\
\hline
\end{tabular}
\label{tab:table1}
\end{table}

\subsection{Data collection}
\label{sec:section2.2}

\subsubsection{Bacterial strains}
\label{sec:section2.2.1}

Six common strains of foodborne pathogenic and spoilage bacteria were used, including three Gram-positive bacterial strains, including \textit{Bacillus coagulans}, \textit{Bacillus subtilis}, \textit{Listeria innocua}, and three Gram-negative bacterial strains, including \textit{E. coli}, \textit{Salmonella} Enteritidis, \textit{Salmonella} Typhimurium. All bacterial strains were stored in tryptic soy broth (TSB, Sigma-Aldrich, St. Louis, MO, USA) supplemented with 15\% v/v glycerol at –80°C. Prior to experimentation, a bacterial glycerol stock was streaked onto a tryptic soy agar (TSA, Sigma-Aldrich) plate and incubated for 24 h. Subsequently, a single bacterial colony was transferred from the TSA plate to 10 mL of TSB, followed by overnight shaking at 175 rpm. With the exception of \textit{Pseudomonas fluorescens}, which was incubated at 30°C, all strains were incubated at 37°C. The fresh overnight culture was then diluted with sterile phosphate-buffered saline (PBS, Fisher Scientific, Pittsburg, CA, USA) to obtain specific concentrations for microcolony cultivation.

\subsubsection{Microcolony cultivation and microscopy}
\label{sec:section2.2.2}

Bacterial microcolonies were formed following our previously published method \citep{Ma2023}. Briefly, 1 mL of bacterial suspension was deposited onto soft TSA plates (0.7\% w/v agarose) and incubated at 37°C for 3 h. To ensure consistency, the thickness of the soft TSA plates was maintained at 1 mm by adding 2 mL of growth media into a 60-mm petri dish. Microscopic images were acquired using an Olympus IX71 inverted microscope, equipped with 20× and 60× objective lenses. For phase contrast imaging, Ph2 objective lenses were used in conjunction with a phase turret to match the phase ring corresponding to the selected objective. For brightfield imaging, the phase turret was adjusted to the open aperture position to allow standard brightfield illumination with a standard objective lens. Raw images were captured using a CCD camera (Model C4742-80-12AG, Hamamatsu, Tokyo, Japan) and the Metamorph imaging software (version 7.7.2.0, Universal Imaging Corporation). All images were acquired as TIF files and subsequently converted into JPG format using the image processing software ImageJ \citep{Schneider2012}. Eeah image had a resolution of 672 × 512 pixels with a pixel size of 107.5 nm.

\subsection{Data preparation and augmentation}
\label{sec:section2.3}

To prepare the datasets for model training, image files were structured to work seamlessly with the PyTorch deep learning framework, leveraging its built-in tools for data loading and augmentation. Images were organized into a hierarchical directory structure, with bacterial species (serving as the true class labels) represented as folder names within parent directories that encoded metadata for laboratory conditions. Image files were loaded using the \texttt{torchvision.datasets.ImageFolder}, and pixel values were normalized to the range of 0–1 by dividing by 255. To maximize data efficiency, augmentation techniques such as flips, random rotations, and random brightness contrast adjustments were applied using the \texttt{albumentations} library \citep{Buslaev2020}. These augmentations were designed to reflect physical variability inherent in microscopic imaging and helped mitigate overfitting by increasing dataset diversity \citep{Simard2003, Olenskyj2022}. 

Datasets were split into training, validation, and test sets based on the specific requirements of source and target domains. Domain adaptation in this study involved training on a larger dataset from a single source domain and testing on smaller datasets from multiple target domains. For the source domain, 15\% of the images for each of the 6 bacterial species were held out as a test dataset, while the remaining images were randomly split into training and validation sets in a 70/30 ratio. This resulted in 377 training images, 162 validation images, and 90 test images. Target domains contained fewer labeled samples compared to the source domain, so a maximum of 5 images per bacterial species was used for training (30 images total). The test sets for the target domains consisted of 90 images for 20×, and 60 images each for BF and 20×–5h domains.

\subsection{Model architecture and training}
\label{sec:section2.4}

\subsubsection{Feature extractor}
\label{sec:section2.4.1}

EfficientNetV2 was selected as the backbone of the proposed architecture due to its superior accuracy and computational efficiency in feature extraction. EfficientNets are a family of CNN architectures with state-of-the-art performance on image classification tasks, offering better accuracy and efficiency compared to other CNN models \citep{Tan2019}. EfficientNetV2, introduced in 2021, further improves training speed and parameter efficiency through a combination of training-aware neural architecture search and progressive scaling \citep{Tan2021}. This progressive learning strategy involves training the model on smaller image size with weak regularization initially (i.e., weak constraints to the learning process), then transitioning to larger image sizes with stronger regularization. This approach ensures efficient model training while improving generalization by capturing both low-level and high-level features. In this study, the EfficientNetV2 backbone served as the shared feature extractor for both classification and domain adaptation tasks, as detailed in the following sections.

\subsubsection{Adversarial domain adaptation}
\label{sec:section2.4.2}

To address domain shifts across optical and biological variability, DANNs were implemented. These networks were designed to achieve domain invariance and accurate classification across multiple target domains while requiring a minimal number of labeled samples. The model was trained on 539 bacterial images under the controlled laboratory conditions in the PC domain and evaluated on target domains with variations in optical setups and microbial sample incubation times, as detailed in Table \ref{tab:table1}. Each target domain had fewer than 5 labeled samples available per bacterial species.

\begin{figure}[h!]
    \centering
    \includegraphics[width=0.9\textwidth]{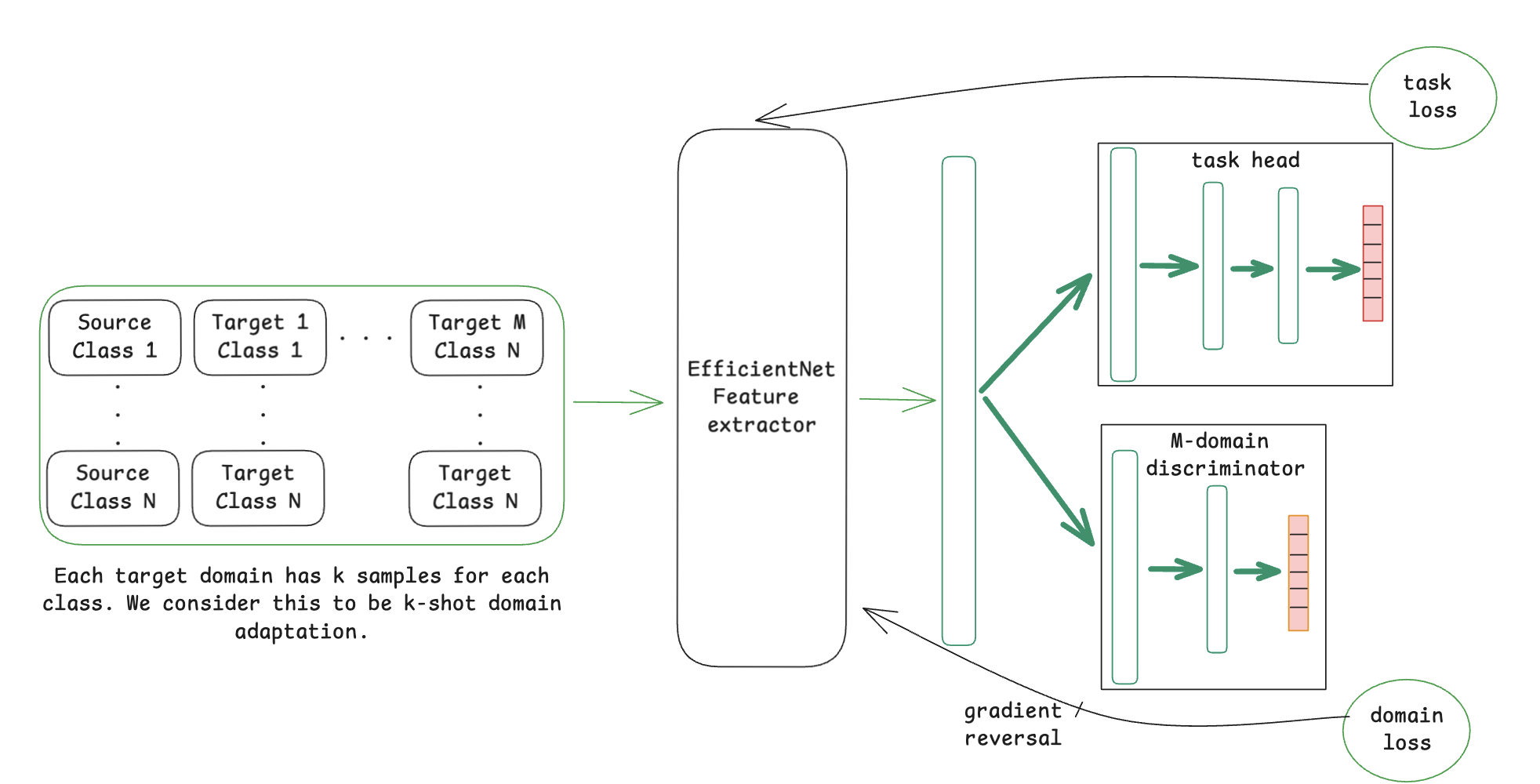}
    \caption{Schematic of the model training process using a domain-adversarial neural network (DANN) for domain adaptation for $M$ domains}
    \label{fig:figure1}
\end{figure}

A variant of DANNs, referred to as MDANN, was also employed to extend domain adaptation to multiple domains simultaneously \citep{Ganin2016}. As shown in Figure \ref{fig:figure1}, the model was designed to learn a shared feature extractor capable of capturing domain-invariant features from microscopic images. These extracted features were then passed to a task-specific classification head to predict class labels (i.e., bacterial species), optimized using a task loss calculated as the cross-entropy between predicted and true class labels. To enforce domain invariance, a multi-domain discriminator network was trained concurrently to predict the domain labels of input images based on their feature representations. The domain loss, calculated as the cross-entropy across predicted domain probabilities (with 0 representing the source domain and $1,\ldots,m$ representing the $m$ target domains), was backpropagated through the discriminator network. A gradient reversal layer was applied to multiply the gradient of weights of the discriminator network with respect to the training loss by $-\lambda$, a tunable hyperparameter that controls the strength of domain alignment. The adjusted gradients were then propagated through the backbone network, promoting the extraction of domain-invariant feature representations. As the discriminator network improved in distinguishing domain labels, the backbone network simultaneously adapted to generate features that minimized domain-specific biases.

\subsubsection{Model training}
\label{sec:section2.4.3}

The model training process was structured to achieve simultaneous classification and domain adaptation using a multi-task learning framework. Let $S$ denote the source domain and $T = \{T_1, \ldots, T_{m-1}\}$ denote the target domains. The combined dataset is denoted as $(S \cup T) = \{S, T_1 \ldots, T_{m-1}\}$, where the class labels for any $x \sim (S \cup T)$ are defined as $\{0, \ldots, 5\}$. The shared EfficientNetV2 feature extractor, $f_e$, is parameterized by $\theta_e$. The forward pass of the extractor is represented as $f_e(x, \theta_e) = e$, where $e \in \mathbb{R}^{2152}$ is the feature embedding of the input $x$ and $\theta_e$ are the learnable parameters of the feature extractor. The extracted features are processed by two separate heads: a classification head and a domain regressor head. The classification head, $f_c(f_e(x, \theta_e),\theta_c): \mathbb{R}^{2152} \to \mathbb{R}^6$, maps the embeddings to probabilities for each class label, while the domain regressor head, $f_d(f_e(x,\theta_e),\theta_d): \mathbb{R}^{2152} \to \mathbb{R}^m$, maps the embeddings to domain labels, where $m$ is the total number of domains (including the  source domain).

The multi-task loss function comprises the classification loss, $\mathcal{L}_C$, and the domain adaptation loss, $\mathcal{L}_D$, and is defined as follows:
\begin{equation}
    \mathcal{L} = \mathcal{L}_C + \mathcal{L}_D
\end{equation}
\begin{equation}
    \mathcal{L}_{C} = -\sum_{i=1}^{m} y_i \cdot log f_c(f_e(x_i, \theta_e), \theta_c)
\end{equation}
\begin{equation}
    \mathcal{L}_{D} = -\lambda \sum_{i=1}^{m} d_i \cdot log f_d(f_e(x_i, \theta_e), \theta_d)
\end{equation}
\noindent where $x_i \sim (S \cup T)$ is an input sample with class label $y_i \in \{0, \ldots, 5\}$ and domain label $d_i \in \{0, \ldots, m-1\}$, where $d_i=0$ corresponds to $S$, $d_i =1$ corresponds to $T_1$, and so on. The hyperparameter $\lambda$ controls the relative importance of the domain loss. A higher value of $\lambda$ increases the alignment of features representations across domains.

Finally, the model parameter are updated during training using the following equations: 
\begin{equation}
    \label{dann_loss}
    \theta_f \leftarrow \theta_f - \alpha \frac{\partial L_C}{\partial \theta_f} + \alpha \cdot \tau \cdot \lambda \frac{\partial L_D}{\partial \theta_f}
\end{equation}
\begin{equation}
    \theta_c \leftarrow \theta_c - \alpha \frac{\partial L_C}{\partial \theta_c}
\end{equation}
\begin{equation}
    \theta_d \leftarrow \theta_d - \alpha\cdot \lambda \frac{\partial L_D}{\partial \theta_d}
\end{equation}
\noindent where $\alpha$ is the learning rate and $p$ is the scaling factor. Gradient reversal is applied in Equation $\ref{dann_loss}$ to the domain-regressor loss in the update of $\theta_f$ by adding rather than subtracting its gradient. The scaling factor $\tau$ is applied to the domain-regressor loss during the gradient reversal step, a monotonically increasing sigmoid function as follows:

\begin{equation}
    \tau = \frac{2}{1 + e^{\frac{-10t_i}{t}}}-1
\end{equation}
\noindent where $t_i$ denotes the current epoch and $t$ is the maximum number of epochs set before training. This adaptive weighting mechanism allows the feature extractor to learn discriminative features in early epochs and transition to domain-invariant features in later epochs, typically after 30-40 epochs \citep{Chen2019}. 

Training was conducted for a maximum of 90 epochs with a batch size of 6. The AdamW optimizer was used, with a learning rate of 0.001 and a weight decay of 0.001. For each experiment, the model checkpoint corresponding to the lowest validation loss was selected for evaluation on the held-out test set. 

\subsection{Model evaluation and visualization}
\label{sec:section2.5}

The performance of bacterial image classification was evaluated using classification accuracy as the primary metric. Models trained on multiple domains were evaluated separately for both the source and target domains to determine their effectiveness in each context.

Additionally, Gradient-based class activation mapping (Grad-CAM) \citep{Selvaraju2017} was employed to gain qualitative insights into the specific image features utilized by the model to predict bacterial species. Grad-CAM is widely used visualization technique in image classification tasks that highlights regions of an input image receiving significant attention from the model during prediction. While the model outputs a single value corresponding to the predicted bacterial species, the Grad-CAM visualizations indicate the regions in the input image that most influence the prediction. In this study, it was hypothesized that these regions would correspond to areas of the input image where individual microcolony is visible, although other features of the input, such as the spatial arrangement of multiple microcolonies, might also contribute to the predictions. 

For qualitative understanding of domain alignment within the model, t-distributed stochastic neighbor embedding (t-SNE) \citep{Hinton2002} was employed to project the feature vectors of test images from the source and target domains into two-dimensional space. This technique enabled visualization of whether domain-adversarial training facilitated domain alignment between the input images of different domains. 

\section{Results}
\label{sec:section3}

\subsection{Domain variability in microcolony imaging}
\label{sec:section3.1}

This study explores domain adaptation across one \textit{source domain} and three \textit{target domains}, each characterized by distinct laboratory conditions (Table \ref{tab:table1}). The source domain (PC) dataset consisted of images collected under controlled laboratory conditions using phase contrast microscopy at 60× magnification from microcolonies incubated for 3 h. Target domains were chosen to simulate variations encountered in real-world applications, including brightfield microscopy at 60× magnification (BF), phase contrast microscopy at a lower magnification (20×), and phase contrast microscopy at 20× magnification with an extended incubation time to capture additional biological variability (20×–5h).

\begin{figure}[h!]
    \centering
    \includegraphics[width=0.9\textwidth]{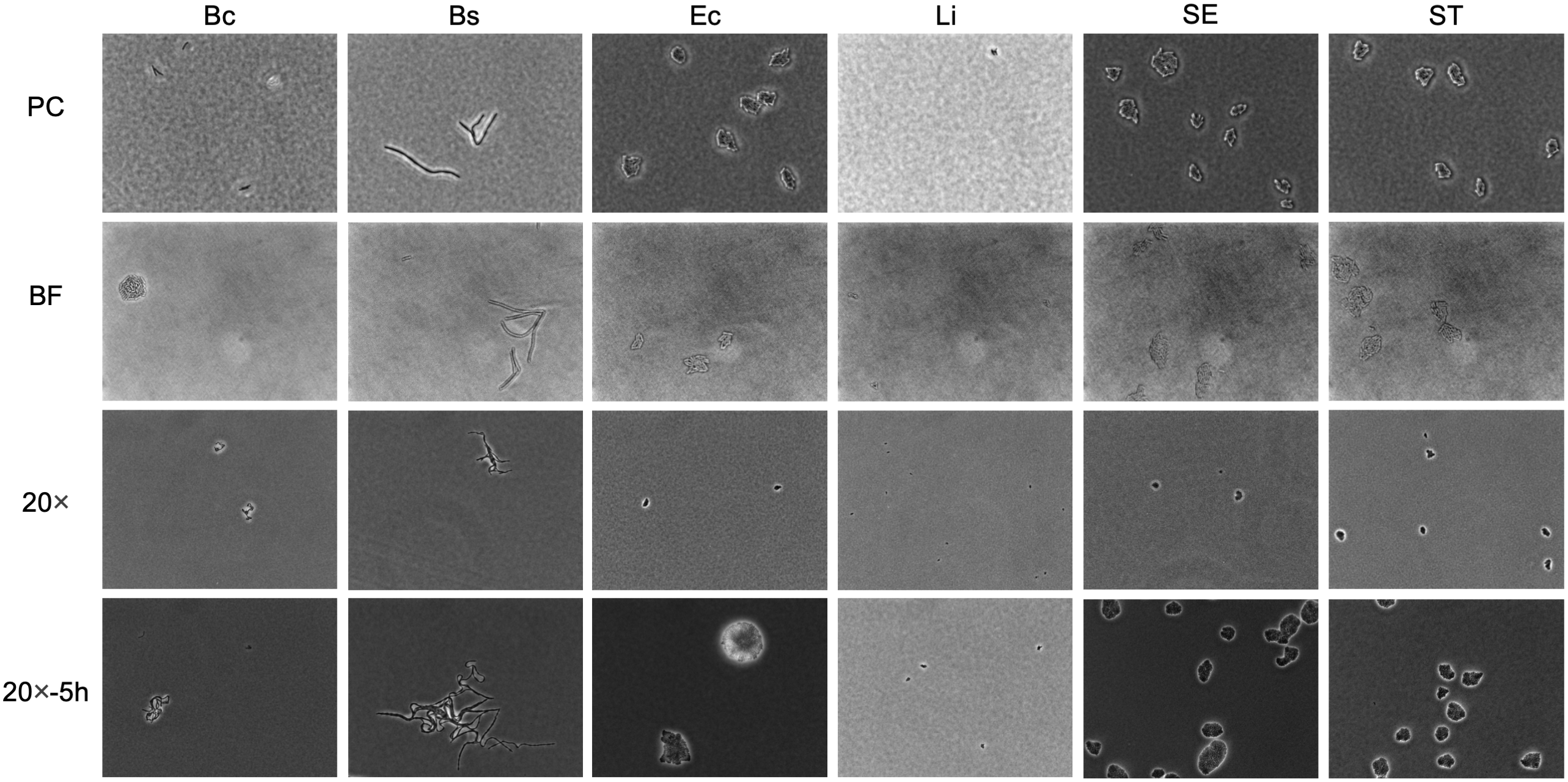}
    \caption{Example of bacterial microcolony images across domains. Columns represent different bacterial species, including \textit{Bacillus coagulans} (Bc), \textit{Bacillus subtilis} (Bs), \textit{Listeria innocua} (Li), \textit{E. coli} (Ec), \textit{Salmonella} Enteritidis (SE), \textit{Salmonella} Typhimurium (ST). Rows represent domains with different laboratory conditions, as detailed in Table 1.}
    \label{fig:figure2}
\end{figure}

Substantial variability was observed in microcolony imaging across domains, as illustrated in Figure \ref{fig:figure2}. Source domain images displayed granular backgrounds and fine distinctions between individual bacterial cells, in addition to well-defined microcolonies. These features were attributable to the higher magnification and phase contrast microscopy under controlled cell growth conditions. In contrast, images from the BF target domain, exhibited lower contrast, rendering microcolonies less discernible against the background. Similarly, images from the 20× domain presented smaller targets with reduced cellular features resolution, attributable to the lower magnification. Extended incubation in the 20×–5h domain resulted in larger microcolony sizes but introduced variability in spatial distributions and subtle focal shifts, further complicating feature discrimination.

This variability in optical and biological factors diminishes the discriminative quality of image features, presenting challenges for bacterial detection and classification. Addressing this variability necessitated models capable of effectively transferring performance from the high-quality source domain (PC) to variable target domains (BF, 20×, 20×–5h). Given that the bacterial species remained consistent across domains, the model needed to learn domain-invariant features that were unaffected by optical and biological variability, while simultaneously maintaining the ability to classify species-specific traits. To achieve this, DANNs and MDANNs were further implemented, as detailed in Section \ref{sec:section2.4}. These architectures and training processes were specifically designed to align feature representations across domains, ensuring both domain invariance and accurate classification across all target domains.

\subsection{Classification performance improvements with domain adaptation}
\label{sec:section3.2}

The ability of DANNs for single-target domain adaptation and MDANNs for multi-domain adaptation  was evaluated to address the classification performance gaps between the source domain (PC) and target domains (BF, 20×, 20×–5h). Single-target domain adaptation aligned feature representations for individual target domains, while multi-domain adaptation generalized across multiple target domains simultaneously. Both approaches leveraged limited labeled samples (1-shot, 3-shot, or 5-shot) from the target domains, enabling significant performance improvements despite substantial variability in optical setups and microbial sample incubation times.

\begin{figure}[h!]
    \centering
    \includegraphics[width=0.8\textwidth]{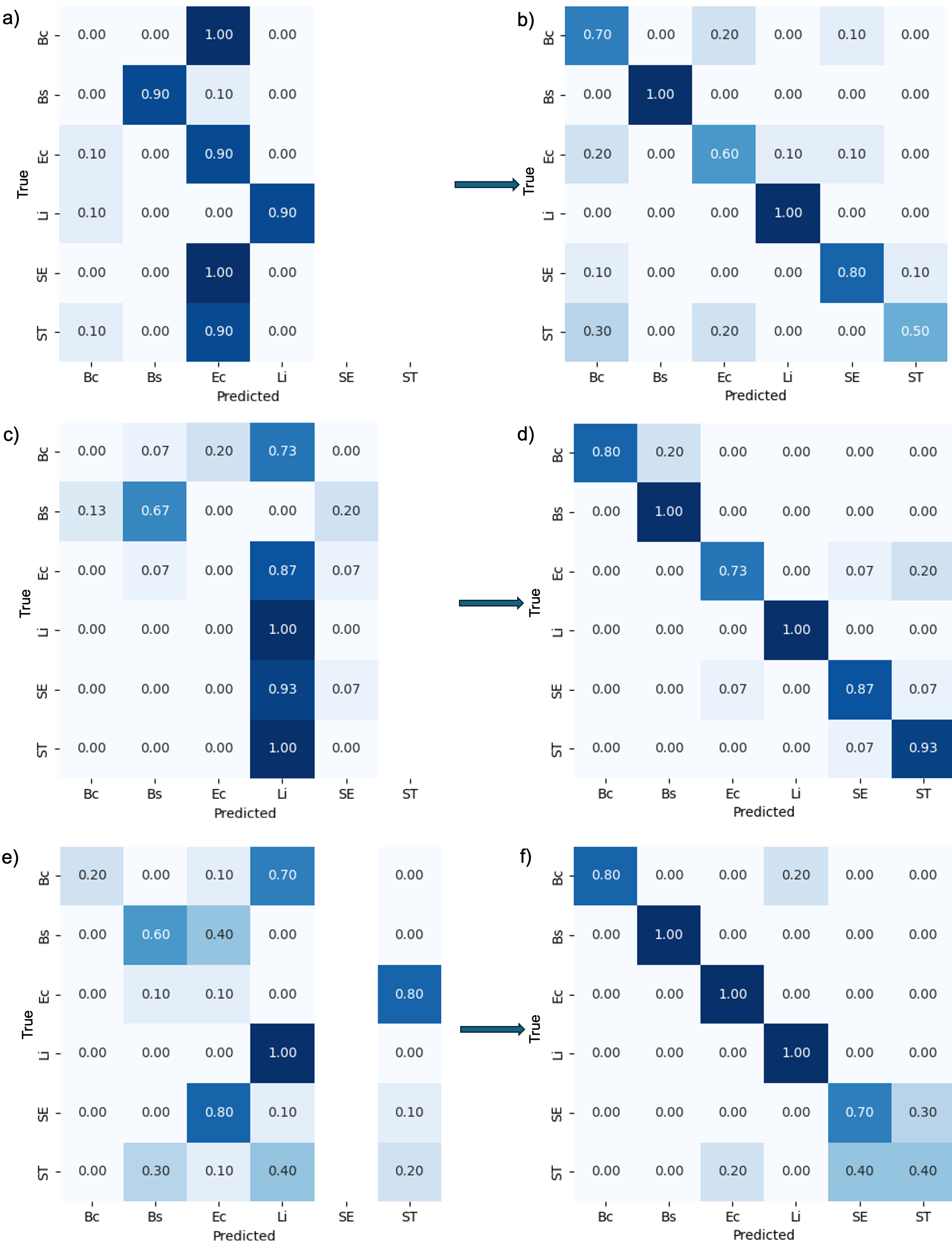}
    \caption{Comparison of classification confusion matrices for source-only training and domain-adversarial training across three target domains. (a, c, e) Classification results for source-only training for the BF, 20×, 20×–5h target domains, respectively. (b, d, f) Improved classification results for the corresponding target domains using domain-adversarial training with DANNs. Bc: \textit{Bacillus coagulans}. Bs: \textit{Bacillus subtilis}. Ec: \textit{E. coli}. Li: \textit{Listeria innocua}. SE: \textit{Salmonella} Enteritidis. ST: \textit{Salmonella} Typhimurium.}
    \label{fig:figure3}
\end{figure}

As shown in Figure \ref{fig:figure3}, DANNs substantially improved classification performance for target domains compared to source-only training. The left panels of Figure \ref{fig:figure3} (a, c, and e) illustrate the limited performance of source-only training, which relied solely on data from the source domain. This approach struggled to generalize to target domains due to domain shifts. In contrast, domain-adversarial training effectively aligned feature representations across domains, enhancing classification accuracy in the target domains, as shown in the right panels (b, d, and f).

The classification performance of single-target domain adaptation using DANNs is quantified in Table \ref{tab:table2}, with target domain accuracy increasing by up to 54.45\%, while source domain accuracy experienced minimal degradation, generally within 4.44\% of source-only training. The model performed better on the 20× and 20×–5h domains compared to BF. Specifically, accuracy improvements reached up to 54.45\% and 43.33\% for 20× and 20×–5h domains, respectively, while BF domain showed a more modest improvements of 31.67\%.

\begin{table}[h]
\centering
\caption{Classification accuracy for domain-adversarial neural networks (DANNs) with EfficientNetV2 backbone. Results are presented for source-only training and domain-adversarial training with 1-shot, 3-shot, 5-shot labeled samples.}
\begin{tabular}{llcccc}
\hline
\multicolumn{2}{l}{} & \textbf{Source-only} & \textbf{1-shot DANN} & \textbf{3-shot DANN} & \textbf{5-shot DANN} \\
\hline
\multirow{2}{*}{BF} & Source & 94.44\% & 91.11\% & 94.44\% & 93.33\% \\
& Target & 43.33\% & 54.44\% & \textbf{75\%} & 73.33\% \\
\hline
\multirow{2}{*}{20×} & Source & 94.44\% & 90.0\% & 90.0\% & 93.33\% \\
& Target & 34.44\% & 54.44\% & 82.22\% & \textbf{88.89\%} \\
\hline
\multirow{2}{*}{20×–5h} & Source & 94.44\% & 93.33\% & 94.44\% & 94.44\% \\
& Target & 40.00\% & 71.67\% & 83.33\% & \textbf{83.33\%} \\
\hline
\end{tabular}
\label{tab:table2}
\end{table}

Multi-domain adaptation using MDANNs further showed strong generalization across multiple target domains, as shown in Table \ref{tab:table3}. MDANNs achieved accuracy comparable or higher than single-target domain adaptation for BF domain, with accuracy improvement of 33.34\% (from 43.33\% to 76.67\%). However, for the 20× domain, MDANNs achieved slightly lower performance compared single-target DANNs, despite showing substantial improvement of 47\% (from 34.44\% to 82.22\%).

\begin{table}[h]
\centering
\caption{Classification accuracy for multi-domain adversarial neural networks (MDANNs) with EfficientNetV2 backbone, where the source domain was PC and the target domains were BF and 20×. Results are presented for domain-adversarial training with 1-shot, 3-shot, 5-shot labeled samples.}
\begin{tabular}{lccc}
\hline
\textbf{} & \textbf{1-shot MDANN} & \textbf{3-shot MDANN} & \textbf{5-shot MDANN} \\
\hline
Source & 87.78\% & \textbf{92.22}\% & \begin{tabular}[c]{@{}c@{}}90\%
\end{tabular} \\
Target (BF) & 60\% & 68.33\% & \begin{tabular}[c]{@{}c@{}}\textbf{76.67}\%
\end{tabular} \\
Target (20×) & 54.44\% & \textbf{82.22}\% & \begin{tabular}[c]{@{}c@{}}75.56\%
\end{tabular} \\
\hline
\end{tabular}
\label{tab:table3}
\end{table}

Overall, these results highlight the effectiveness of domain-adversarial training in both single-target and multi-domain scenarios. DANNs and MDANNs demonstrated the ability to learn domain-invariant features and align feature representations across domains, enabling accurate bacterial classification despite the variations in optical setups and microbial sample incubation times.

\subsection{Visualizing feature representations and domain alignment}
\label{sec:section3.3}

\begin{figure}[h!] 
    \centering
    \begin{subfigure}{0.3\linewidth} 
        \includegraphics[width=\linewidth]{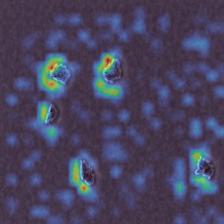} 
        \caption{PC domain}
        \label{fig:figure4a}
    \end{subfigure}
    \hfill
    \begin{subfigure}{0.3\linewidth} 
        \includegraphics[width=\linewidth]{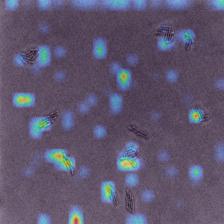}
        \caption{BF domain}
        \label{fig:figure4b}
    \end{subfigure}
    \hfill
    \begin{subfigure}{0.3\linewidth} 
        \includegraphics[width=\linewidth]{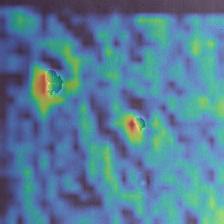} 
        \caption{20× domain}
        \label{fig:figure4c}
    \end{subfigure}
    \caption{Gradient-based class activation mapping (Grad-CAM) visualizations of feature representations for a 5-shot MDANN. Activations are shown for the source domain (a: PC) and target domains (b: BF, c: 20×), with warmer colors indicating stronger model attention.}
    \label{fig:figure4}
\end{figure}

Grad-CAM visualizations (Figure \ref{fig:figure4}) provided insights into how the model identified important features across different imaging domains, highlighting the regions of microcolony images that were most influential during classification. Warmer colors (e.g., red and yellow) represented areas with stronger activation, indicating the model’s focus during prediction. In the PC domain, activations were sharply localized along the well-defined edges of individual microcolonies, reflecting the high-contrast details available in phase contrast microscopy under controlled laboratory conditions. In the BF domain, activations were broader and slightly less concentrated. While the overall activation structure resembled that of the PC domain, the model adapted to the lower contrast of brightfield microscopy by expanding its focus to include central regions of microcolonies. In the 20× domain, activations were noticeably more diffuse and less sharply defined. The lower magnification reduced the resolution of individual cells, making edge-specific features less informative. As a result, the model shifted its attention to the spatial arrangement and overall morphology of microcolonies, which became the most informative distinguishing features.

\begin{figure}[h!]
    \centering
    \includegraphics[width=0.5\textwidth]{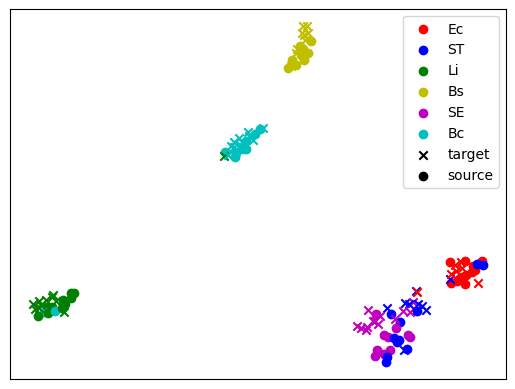}
    \caption{t-distributed stochastic neighbors embedding (t-SNE) visualization of feature embeddings extracted by the 5-shot DANN. The source domain is PC, and the target domain is 20×–5h. Clusters represent feature embeddings for specific bacterial species, with source samples (black dots) and target samples (colored crosses). Overlapping clusters indicate effective alignment of features across domains. Bc: \textit{Bacillus coagulans}. Bs: \textit{Bacillus subtilis}. Ec: \textit{E. coli}. Li: \textit{Listeria innocua}. SE: \textit{Salmonella} Enteritidis. ST: \textit{Salmonella} Typhimurium.}
    \label{fig:figure5}
\end{figure}

The alignment of feature representations was further validated using t-SNE visualization (Figure \ref{fig:figure5}). This visualization projected high-dimensional feature embeddings into a two-dimensional space, allowing the formation of visually distinct clusters for each bacterial species. In the visualization, the source domain (PC) embeddings are represented by circles and the target domain (e.g., 20×–5h) embeddings are represented by crosses, with colors indicating different bacterial species. Overlap or close alignment between source and target embeddings within the same bacterial species cluster indicates that domain-adversarial training successfully aligned features across domains. For most species, such as \textit{Bacillus coagulans} (Bc), \textit{Bacillus subtilis} (Bs), \textit{E. coli} (Ec), and \textit{Listeria innocua} (Li), the source and target clusters overlap significantly, demonstrating effective alignment of features across domains. However, for closely related species such as \textit{Salmonella} Enteritidis (SE) and \textit{Salmonella} Typhimurium (ST), the visualization shows overlapping clusters with minimal separability between these two species, revealing challenges in distinguishing them.

\section{Discussion}
\label{sec:section4}

\subsection{Generalizing bacterial classification beyond sample preparation and long incubation constraints}
\label{sec:section4.1}

This study demonstrates the application of adversarial domain adaptation to enhance the generalizability of AI-enabled microscopy for bacterial classification under diverse laboratory conditions. By integrating a pretrained EfficientNetV2 backbone with domain-adversarial training, our approach achieved up to a 54.45\% improvement in classification accuracy across target domains, effectively addressing domain shifts such as variations in optical setups and microbial sample incubation times. Previous studies in AI-enabled bacterial classification predominantly focused on optimizing model performance within controlled laboratory settings, often relying on sample preparation techniques such as staining, selective enrichment, or the use of specialized reagents, which extended processing times and required skilled personnel \citep{Chen2024, Ramesh2024, Wakabayashi2024}. For example, \cite{Chen2024} utilized selective media and Gram staining to isolate and chemically label bacterial samples, enabling their model to classify pre-processed colonies but constraining the scope to well-prepared datasets, with incubation times of 18–24 h for enrichment and an additional 18–48 h for further processing. Similarly, many studies, including \cite{Ramesh2024}, utilized the publicly available Digital Images of Bacteria Species (DIBaS) dataset, which consisted of Gram-stained samples imaged at high magnification (100×), providing high-quality data but without addressing generalizability across broader sample preparation and imaging conditions. \cite{Wakabayashi2024} introduced variability by working with inoculated meat samples, yet relied on extended incubation times (20 h) and selective media, which may not be feasible in resource-limited settings. In these studies, AI models were employed to analyze highly processed or enriched datasets, often focusing on tasks like counting fully grown colonies or identifying pre-processed features rather than generalizing to diverse conditions. In contrast, our study focused on minimizing the reliance on extensive sample preparation, long incubation times, and precise skillsets required for high-quality image acquisition. By using general-purpose media with a reduced incubation time of 3 h, we demonstrated  the potential to shift bacterial classification from labor-intensive preparation to automated abstract feature recognition. To the best of our knowledge, this is the first study to explicitly evaluate the generalizability of a model trained on high-quality, controlled datasets to diverse image domains, encompassing optical and biological variability. These advancements highlight the model’s potential for resource-limited and variable laboratory environments.

\subsection{Adversarial domain adaptation for cross-modality imaging in bacterial classification}
\label{sec:section4.2}

The results of this study highlight the effectiveness of adversarial domain adaptation for bacterial classification in microscopy, showcasing its ability to address optical variability and reduce dependence on specific microscopy modalities and resource-intensive sample preparation. Cross-modality microscopy has been extensively explored in biomedical imaging, particularly for tasks like cell and nucleus detection, where large, curated datasets are often used to achieve robust performance. For example, previous work utilized k-means clustering and modality-specific deep learning models for cell segmentation across diverse microscopy modalities, such as brightfield, phase contrast, differential interference contrast (DIC), and fluorescence microscopy, emphasizing the need for domain adaptation to improve generalizability beyond individual models \citep{Wang2023}. Similarly, adversarial domain adaptation was applied to translate between stained and unstained images for cell and nucleus detection, improving detection performance compared to baseline "source-only" training \citep{Xing2021}. However, these studies primarily focused on instance detection and often relied on staining information for classification. Another study applied domain adaptation to digitally stain white blood cells, reducing the need for staining procedures and preserving structural information critical for classification, though it did not address optical variability, such as changes in magnification \citep{Tomczak2021}. In contrast, this study emphasizes optical variability for bacterial classification, exploring the practicality of adversarial domain adaptation with small datasets containing as few as 1–5 labeled samples per species in target domains. It demonstrated robust performance across variations in optical setups, including transitions between microscopy modalities (phase contrast to brightfield) and magnifications (60× to 20×). This few-shot approach can be particularly advantageous in scenarios where collecting and annotating large datasets is impractical. By reducing dependency on extensive labeled data, the proposed framework ensures scalability for real-world applications. A key advantage of this approach lies in its ability to learn abstract features that are domain-invariant, enabling robust classification of both Gram-positive and Gram-negative bacteria across diverse imaging conditions without relying on staining or contrast-enhancing techniques. Unlike earlier bacterial classification methods that required Gram-staining to enhance image contrast \citep{Chen2024, Ramesh2024} or phase contrast microscopy for microcolony visualization \citep{Ma2023}, this method shifts the classification burden to the model, bypassing optical and chemical dependencies inherent to traditional workflows. Additionally, transitioning from 60× to 20× magnification facilitates the use of more accessible, less specialized equipment, balancing resolution with broader applicability. By leveraging a few-shot domain adaptation approach to address variability in optical setups, this domain-adversarial framework ensures scalability while maintaining robust performance. This makes it particularly suitable for decentralized and resource-limited environments such as onsite food safety testing.

\subsection{Learning abstract features for small targets and biological variability}
\label{sec:section4.3}

The EfficientNetV2 backbone showed remarkable effectiveness in this study, particularly for classifying small bacterial microcolonies under challenging imaging conditions, such as lower magnifications in the 20× domain. Its compound scaling strategy, which balances depth, width, and resolution, enables hierarchical feature learning that is well-suited for capturing fine-grained features of small targets. Unlike previous approaches that employed shallow CNN architectures, such as the five-layer convolutional model used for foodborne pathogen classification \citep{Chen2024}, EfficientNetV2’s deeper architecture (13–60 convolutional layers) excelled in extracting information-rich abstract features. While \cite{Chen2024} achieved promising results for rapid detection using brightfield microscopy, their approach relied on Gram-staining, high magnifications with oil immersion objectives, and extended incubation times to grow fully formed colonies. In contrast, our EfficientNetV2 backbone, combined with adversarial domain adaptation, enabled classification of microcolonies cultivated for shorter incubation periods without staining or high-resolution imaging. Additionally, previous approaches in bacterial image classification often excluded low-quality images manually to mitigate errors \citep{Chen2024}, whereas EfficientNetV2’s ability to generalize stemmed from its progressive learning strategies, advanced regularization techniques, and integration with an image augmentation pipeline. These capabilities allowed the model to perform robustly even with suboptimal image quality. EfficientNetV2 also addressed limitations of object detection models like YOLOv3, which struggled with small bacterial targets due to specific design constraints \citep{Wakabayashi2024}. Its grid cell mechanism often failed to localize small microcolonies accurately, downsampling erases fine-grained details, and predefined anchor boxes were poorly suited to microcolony sizes and shapes. These limitations, observed in our preliminary experiments with YOLOv4, hindered performance in the 20× domain. In contrast, EfficientNetV2’s hierarchical feature learning and dynamic scaling captured subtle spatial and morphological details, making it well-suited for small bacterial target analysis under varying conditions.

The flexibility of EfficientNetV2 is further supported by Grad-CAM and t-SNE analyses, which highlighted its ability to adapt to varying imaging conditions. Grad-CAM visualizations revealed that in the PC domain, the model focused on edge-specific features, leveraging high-contrast details provided by phase contrast microscopy. For the 20× domain, where reduced magnification limited the visibility of fine cellular features, the model shifted its focus to broader microcolony-level characteristics such as shape, size, and spatial arrangement, demonstrating dynamic feature extraction. Similarly, the t-SNE visualization for the 20×–5h domain showed significant alignment between source (PC) and target embeddings for most species, confirming the model’s ability to learn domain-invariant features and generalize across optical and biological variability. However, overlapping t-SNE clusters for closely related species, such as \textit{Salmonella} Enteritidis (SE) and \textit{Salmonella} Typhimurium (ST), underscored the challenge of distinguishing morphologically similar species, as also reflected in the confusion matrix (Figure \ref{fig:figure3}f). These findings emphasized the combined strength of EfficientNetV2 and adversarial domain adaptation in handling variability across domains.

\subsection{Future directions for enhancing domain adaptation in microscopy}
\label{sec:section4.4}

This study highlights the strengths and challenges of DANNs and MDANNs in addressing domain variability across diverse optical and biological conditions. While DANN improved classification performance across multiple target domains, results from Table \ref{tab:table2} and Section \ref{sec:section3.2} indicate that low contrast in the BF domain posed greater alignment difficulties than variations in magnification or microbial sample incubation times in other domains. MDANNs demonstrated strong generalization across multiple domains by leveraging shared features across bacterial species, but results from Table \ref{tab:table3} and Section \ref{sec:section3.2} show that the BF domain consistently exhibited lower improvements compared to others, underscoring the persistent challenges of low-contrast imaging. Building on prior multi-domain adversarial approaches, which have demonstrated success with larger datasets and complex discriminator architectures \citep{Zhao2018, Pei2018}, this study extended these advancements by achieving domain-adversarial training with as few as 1–5 labeled samples per bacterial species. This few-shot approach reduces the labor-intensive process of manual annotation and aligns with the broader CNN-based strategies prevalent in this field. However, a promising future direction would be to transition toward unsupervised domain adaptation, eliminating the reliance on labeled samples entirely. Leveraging generative adversarial networks (GANs), we could digitally generate pixel-level labels or synthetic training data, enabling an unsupervised workflow that emphasizes pixel-level classification. This shift would address both labor limitations and the need for enhanced alignment at finer spatial resolutions, such as in low-contrast modalities like brightfield microscopy. GAN-based approaches for digital staining or synthetic dataset generation have shown potential in recent studies on biomedical imaging \citep{Mukherjee2023, Goyal2024}, and their integration could further improve domain alignment and classification performance. Additionally, future efforts could incorporate spectral or biochemical features into the dataset, improving the differentiation of closely related bacterial species, such as \textit{Salmonella} Enteritidis and \textit{Salmonella} Typhimurium, providing additional discriminatory power. These advancements would enhance the robustness of domain-adversarial frameworks and broaden their scalability and applicability to decentralized or resource-limited laboratory environments.

\section{Conclusions}

This study demonstrates the potential of adversarial domain adaptation for bacterial classification in microscopy, addressing variability in optical setups and microbial sample incubation times. By integrating EfficientNetV2 image classification backbone with DANNs and MDANNs, the framework achieved robust generalization from controlled laboratory conditions to various setups with optical and biological variability. Single-target adaptation improved performance across individual domains, while multi-domain adaptation achieved strong generalization across multiple domains with slight trade-offs in some cases. Grad-CAM and t-SNE visualizations validated the model’s ability to extract domain-invariant features, enabling robust adaptation to challenges such as low-contrast imaging, low magnifications, and extended incubation times. The few-shot learning approach underscores the scalability of this framework for real-world applications, such as onsite food safety testing, where labeled data are limited. Follow-on studies could explore unsupervised approaches to further enhance domain adaptation by reducing reliance on labeled samples, improving contrast in low-quality images, and enabling alignment at finer spatial resolutions. This study establishes a strong foundation for optimizing multi-domain adaptation and enhancing scalability and accuracy in decentralized and resource-limited settings.

\section*{Acknowledgments}

This work was supported by the Michigan State University startup funds, as well as the USDA-National Institute of Food and Agriculture (grant 2021-67021-34256) and USDA/NSF AI Institute for Next Generation Food Systems (grant 2020-67021-32855).

\bibliographystyle{plain}

\end{document}